# Predicting Patient Risk of Readmission with Frailty Models in the Department of Veteran Affairs

Saeede Ajorlou, Issac Shams, and Kai Yang

*Abstract*— Reducing potentially preventable readmissions has been identified as an important issue for decreasing Medicare costs and improving quality of care provided by hospitals. Based on previous research by medical professionals, preventable readmissions are caused by such factors as flawed patient discharging process, inadequate follow-ups after discharging, and noncompliance of patients on discharging and follow up instructions. It is also found that the risk of preventable readmission also may relate to some patient's characteristics, such as age, health condition, diagnosis, and even treatment specialty. In this study, using both general demographic information and individual past history of readmission records, we develop a risk prediction model based on hierarchical nonlinear mixed effect framework to extract significant prognostic factors associated with patient risk of 30-day readmission. The effectiveness of our proposed approach is validated based on a real dataset from four VA facilities in the State of Michigan. Simultaneously explaining both patient and population based variations of readmission process, such an accurate model can be used to recognize patients with high likelihood of discharging non-compliances, and then targeted post-care actions can be designed to reduce further rehospitalization.

## I. INTRODUCTION

Hospital readmission rates have been identified as a main measure of quality of care received by patients [1] since they are happened due to such factors as premature discharging process or inadequate access to care. More importantly, it is found that rehospitalization causes an unfitting share of costs for inpatient hospital cares. In 2009, [2] reported that 19.6% of Medicare fee-for-service patients discharged from a hospital were readmitted within 30 days, 34.0% within 90 days, and more than half (56.1%) within one year of discharge, collectively accounting for $15 billion of Medicare spending. And recently, based on Obama Care Rule (known as Patient Protection and Affordable Care Act or PPACA), about two-thirds (or 2,211) of U.S. hospitals have been penalized a cumulative $280 million (1%) in Medicare funds because of excess readmissions starting Oct. 1, 2012. This cut will grow to maximum of 2% for the 2014 program year and 3% for 2015 [3]. In this paper, we propose a risk prediction model based on hierarchical nonlinear mixed effect framework to extract significant prognostic factors associated with patient readmissions that mainly caused by patient non-compliance to the medication instructions. The novelty of our method is to directly incorporating patients' history of readmissions, along with other patient characteristics, into the modeling framework thus enabling one to explain both patient and population based variations of readmission process at the same time.

The rest of this paper is arranged as follows. The research methodology with detailed inference and optimization is outlined in section 2. Section 3 explains the data structure and steps to implement the proposed algorithm. Conclusions and future research directions are presented in section 4.

## II. METHODOLOGY

### A. Preliminaries

We define T as a random point in time that a patient is readmitted to the hospital, or is censored from the study by any reason, where $0 < T < 30$. In this study we focus only on readmissions caused by patient non-compliance to the medication instructions. Such data frame was obtained after careful screening done by VA Center for Applied Systems Engineering to narrow the possible causes of re-admission to immature post-care practices. The study begins whenever a patient is discharged from a facility and ends at the 30th day, setting $\sigma = 1$ if the patient was readmitted at either of VA hospitals within the interval and $\sigma = 0$ otherwise. In case of no rehospitalization during the 30-day interval, the event indicator $\sigma$ becomes zero and the readmission time random variable T gets 31. Such observations are called right-censored, which form the only type of censoring in our study. We assume that censoring is non-informative due to the fact that the censoring random variable U is the arranged end of the study and does not have any information about the distribution of T.

At this point, two different modeling approaches with distinct objectives may be applied. The first one is a set of linear (non-linear) classification methods, which focus on re-admission indicator $\sigma$ and try to predict it with inputted attributes like naïve Bayes classifiers and logistic regression. The second one consists of event history analytical models, which focus on random readmission time T and try to study the stochastic behavior therein. In the current study, we decide to go with the second approaches due to the facts that classification-based models (1) mostly cannot be able to predict multiple readmission times of a patient with taking into account the inter-dependencies among them; (2) cannot deal with censoring times and their effects on risk of readmission; and (3) cannot handle time-dependent attributes

*Resrach supported by Veteran Engineering Resource Center-VA-CASE(2011-2012).

Issac Shams is with the Department of Industrial and Systems Engineering, Wayne State University, Detroit, MI, 48202 USA. (e-mail: er7671@wayne.edu).

Saeede Ajorlou, is with the Department of Industrial and Systems Engineering, Wayne State University, Detroit, MI, 48202 USA. (e-mail: er7212@wayne.edu).

Kai Yang, is with the Department of Industrial and Systems Engineering, Wayne State University, Detroit, MI, 48202 USA. (e-mail: ac4505@wayne.edu).

whose values might change during follow-up period after patient discharge.

One suitable way to describe a readmission distribution is through hazard function, which can be defined as an instantaneous risk at which events (readmissions) occur, given no previous events, i.e.

$$h(t) = \lim_{\Delta t \to \infty} \frac{pr(t < T \le t + \Delta t \mid T \ge t)}{\Delta t} \quad (1)$$

In interpreting the hazard function, it is worth to note that the hazard can only be thought as a characteristic of patients not of populations or samples, that is each patient may have a totally different hazard comparing to another patient. There have been lots of applications for different hazard functions, which can incorporate increasing, decreasing, or bathtub shapes in such situations as survival after ER surgery and age-specific mortality studies [4]. The simplest form occurs when the hazard is *constant* over time, i.e., $h(t) = \lambda$, or equivalently, $\log h(t) = \mu$, which implies an exponential distribution for the time till an event takes place, such as survival of patients with advanced chronic disease. All of these models can readily be extended to allow for the effects of covariates with fixed or varying measurements over time. For developing a predictive model in which the risk of an event depends on (fixed or time-dependent) covariates, there are two broad classes of regression models in the literature, namely, Proportional Hazard (PH) models and Accelerated Failure Time (AFT) models. In PH models the hazard is governed by $h(t; Z) = h_0(t) e^{\beta Z}$, where $h_0(t)$ is the baseline hazard function illustrating how the risk of an event changes over time at the baseline levels of covariates (Z=0) whilst β is the coefficient set of prognostic factors illustrating the influences of covariates on hazard variations. The main assumption here is that the ratio of hazards for any two individuals is unchanged over time so we can estimate the parameters without having to make any assumptions about the form of $h_0(t)$. In AFT models, in contrast, the effect of a covariate is expressed by multiplication of *event time*, not hazard, by some constant. This type of modeling can be framed as parametric linear models for the natural logarithm of $T$, i.e., $\log(T) = \mu + \beta Z + \sigma \varepsilon$ where ε is a random error term with some parametric distributional assumptions and σ is the variance of disturbance [5].

Although the semi-parametric PH models along with multivariate logistic regression, have been used by some authors for studying the readmission process [6], a few support the significance of patients' history of readmission on predicting the future risk of readmission. Besides, they have not directly incorporated this effect into their modeling framework to control the variations of repeated rehospitalization and its marginal influence on each patient's admission/discharge profile [7,8]. To overcome this shortcoming, in the next part, we explain our proposed approach, which also allows additional clinical and demographic covariates being included into modeling structure.

*B. Modeling Framework*

Generally there are two conventional approaches dealing with repeated event studies. One is to perform a separate analysis for each successive event (readmission) and then make an overall interpretation of effect contributions, which may be tedious and leads to lots of ambiguities for event-to-event variations. The other is to pooling all the events together, treating each as a distinct observation, and estimate a single model, which introduces the problem of having dependencies among multiple events and consequently attenuates the estimates of covariate effects. Therefore, an approach is of interest to explicitly control for dependencies among successive readmissions and at the same time, correct for biases in estimates and test statistics [9]. Such models are often called as *subject-specific* in survival analysis.

Letting $\lambda_{ij}(t)$ be the risk of $j^{th}$ readmission for patient i ($j = 1, 2, \ldots, n_i$) at time $t$, we may propose

$$\lambda_{ij}(t) = \lambda_0(t) e^{\beta_{ij}(t) + \gamma_i} \quad (2)$$

where $\lambda_0(t)$ is an arbitrary baseline hazard rate, $Z_{ij}$ is the collected vector of (*fixed-effect*) associated risk factors, β is the vector of unknown coefficients, and $\gamma_i$ is the *random effect* for patient profile $i$. It should be noted that the random components $\gamma_i$ is subscripted by $i$ but not by j, declaring that the random (unobserved) term is constant from one readmission to the next. Additionally, they are assumed to be independent and identically distributed with an unknown but fixed dispersion parameter, and the distribution of $\gamma_i$ is independent of $Z_{ij}$. Here, we describe some specifications that make our method versatile for addressing the hospital readmission process. Firstly, the response (readmission time) evolves over time *within* patients from a cohort of interest. Secondly, there is a continuous mechanism that underlies patient profiles of repeated measurements of the response to vary in the patient population. Thirdly, inter- and intra-patient variability of readmission process has to simultaneously be elucidated [10]. Lastly, there should be some space for incorporation of both fixed and random patient effects without any pre-specified form.

*C. Optimization*

Generally, implementing HNLM models involves iterative numerical optimizations, which is a potentially high dimensional and computationally expensive thanks to intractable integrations and analytical approximations and heuristics methods [11,12]. Different approaches, by the way, have been proposed and applied in the literature that may be classified into three main groups. The first two, *First-order methods* and *First-order conditional methods*, employ Taylor series (or Laplace's) approximation of $y_i$ (or conditional expectation $E(y_i|u_i)$ about a moment of random effect and solves a set of generalized estimating equation (GEE) based on those marginal moments. The last approach,

which we take on in this study, is "Exact likelihood method" that directly uses deterministic (or stochastic) approximation to the integrals. Particularly, we use adaptive Gauss-Hermite quadrature (deterministic) as described in [13] which centers the integral at the empirical Bayes estimate of $u_i = argmin\{-log[p(y_i|Z_i, \varphi, u_i)q(u_i|\xi)]\}$ with $\varphi$ and $\xi$ set equal to their current estimates. In a nutshell the proposed algorithm selects the number of quadrature points adaptively by evaluating the log-likelihood function at the initial values of the unknown parameters until two successive iterations have a relative difference less than 1E-4. To carry out the minimization problem of $f(\theta)$, a number of nonlinear optimization techniques have been suggested by some authors and put into practice in various circumstances [14,15]. Also due to the fact that no single algorithm exists that always reaches the global optimum (in a reasonable amount of time), no superiority is realized among them. Yet some schemes call for less time and memory since they only compute, in each iteration, the function value (optimization criterion) and the gradient vector, but not the Hessian matrix. To mention a few, the double-dogleg method, conjugate gradient methods, and quasi-Newton methods are popular examples of this type. In this paper, the dual quasi-Newton method is chosen since it demonstrates more balance and stability comparing to its competitors and requires only the gradient to update an approximate Hessian matrix (based on finite-difference approximate of derivatives), which would be promising and time-saving for medium to moderately large optimization problems [16]. We, in addition, set the convergence criterion based on maximum absolute gradient $\max_j |g_j(\theta^{(k)})| \leq$ 1E-8 where $\theta^{(k)}$ refers to the (unknown) parameter vector at the $k^{th}$ iteration and vector $g(\theta)$ refers to the gradient vector $\nabla f(\theta)$. In summary, in each iteration $k$, the dual quasi-Newton method performs some iterative line-search algorithms seeking to optimize a linear approximation of nonlinear objective $f(\theta)$ along a feasible descent direction $l^{(k)}$,

$$\theta^{(K+1)} = \theta^{(K)} + \eta^{(k)} \ell^{(k)} \quad (3)$$

by computing a nearly optimal scalar $\eta^{(k)}$.

III. DATA AND ANALYSIS

We are given a dataset consisted of four VA medical center (VAMC) facilities, namely, Ann Arbor, Battle Creek, Detroit, and Saginaw in the state of Michigan during first halves of 2008 up to 2012. There are 6685 randomly selected records corresponding to 5180 different admitted inpatients along with 37 covariates. During this period of time, patients may be moved to different wards within a hospital for receiving different cares and thus create various admit/discharge profiles. All patient factors, except date of birth, admission date, and discharge date, are measured in nominal scale. To achieve a better picture of data environment, we tentatively arrange attributes into five categories: demographic (marital status, race, etc.), financial (insurance status, employment status, etc.), war-connected (prisoner-of-war status, radiation status, etc.), admission-connected (ward, enrollment priority), and healthcare-connected (diagnosis-related group, ICD-9, etc.). In next parts we take practical steps to analyze the data with our proposed approach.

*A. Application to VAMC patients*

To appropriately apply our proposed approach, we first need to select a candidate set of fixed-effect covariate $Z$ (the random-effect is taken care with feature "id") and set the starting values of parameter estimates $\theta^{(0)}$. Then the dual quasi-Newton technique discussed in part (2.4) is called to iteratively optimize the vector of estimates until the convergence criterion is met. This is done by package 'optimx' in R as described in [17]. Since values of $\theta^{(0)}$ can considerably affect the convergence and the computation time of the algorithm, we decide to set fixed-effects estimates to zero (but intercept to one), scale parameter $\omega$ to zero (corresponding to exponential baseline hazard), and $\sigma_u^2$ to one. Alternatively, one can obtain good initial coefficient estimates from fitting general Cox PH model. For choosing the candidate vectors of $Z$, we investigate within- and between-group correlations of attributes to realize the structure of dependencies and avoid for problems of multicollinearity and over-specification. As an instance, the variables "Vietnam Status" and "Radiation Status" from war-connected category are highly correlated and, at the same time, "Employment Status" from financial category is correlated with "Admission Eligibility" from admission-connected category. In addition, following reference cell coding scheme, several dummy variables are generated for preparing a non-singular design matrix of categorical attributes.

As another important point to be considered, we effectively explore the appropriate functional forms of the features to be included in regression relation by plotting the (function of) response against the (function of) feature. For example, LOS is appeared to have a logarithmic relation with log response (i.e., $log \lambda_{ij}(t)$). Ultimately, stepwise selection procedure (with P-Value to enter and remove of 0.1 and 0.15, respectively) is applied while considering different types of effects, such as interaction or nested effects. The selected effects are (I) admission source with 2 levels VA hospital and NHCU (Nursing Home Care Unit) as the reference level (II) patient sequence (III) patient LOS (IV) marriage status with 3 levels 'married', 'previously married', and 'never married' as the reference category (V) user enrollment status with 2 levels 'YES' and 'NO' as the reference class. We also examined Schwarz's criterion (also known as BIC) to the dataset and found that the model, which minimizes this criterion, again contain the above features [18, 19].

The converged coefficient estimates and is summarized in Table 1. In these tables "$b_0$" is the intercept, "badm" stands for admission source coefficient, "bseqadm" is the coefficient for {sequence*admission source} interaction, "bseqloglos" denotes coefficient for {sequence*log (LOS)} interaction, "b1maruser" is the coefficient for {marriage status*user enrollment status} for those who were married and enrolled (in VAMC) for the following fiscal year, "b2maruser" is the same as "b1maruser" but for those who were previously married and enrolled for the following fiscal year, "sd" denotes the standard error of the random effect, and "omega" is the Weibull scale parameter.

Table 1: Parameter Estimates

| Parameter | Estimate | St. Error | P-Value | 95% Confidence Interval |
|---|---|---|---|---|
| b0 | -8.7573 | 0.4963 | <.0001 | (-9.7304, -7.7841) |
| badm | 4.7175 | 0.4359 | <.0001 | (3.8628, 5.5723) |
| bseqadm | -1.1979 | 0.1176 | <.0001 | (-1.4284, -0.9673) |
| bseqloglos | -0.1795 | 0.0533 | 0.0008 | (-0.284, -0.0750) |
| b1maruser | -0.1581 | 0.1464 | 0.2805 | (-0.4452, 0.1291) |
| b2maruser | 0.3321 | 0.1229 | 0.0069 | (0.0911, 0.5730) |
| sd | 2.7614 | 0.1794 | <.0001 | (2.4096, 3.1132) |
| omega | 0.0003 | 0.1488 | <.0001 | (0.0000, 0.0005) |

The key results are highlighted here. The random-effect standard error is large and highly significant which verifies that there is certainly unobserved heterogeneity (dispersion) across patients (or there is dependence among the repeated readmissions) and (ii) the scale parameter greatly turns out to be zero which corresponds to exponential baseline hazard. Except "b1maruser" all other features are significantly contributed to the patient risk of readmission. An example of coefficient interpretation may be expressed like: 'the hazard of readmission, controlling for other covariates, for those admitted in hospital is near 4.72% of the hazard for those admitted in NHCU'.

For measuring the goodness of fit of our model, we calculate the generalized $R^2$ as

$$R^2 = 1 - \left(\frac{L(0)}{L(\hat{\theta})}\right)^{2/n} \quad (4)$$

which ends up to be 79.68% in our study. In other words, near 80% of all variation in patient readmission risk can be explained by our method. In the above relation, $L(0)$ is the likelihood of the model with only the intercept, $L(\hat{\theta})$ is the likelihood of the estimated model, and $n$ is the sample size. Also one may be interested in testing the overall contribution of {marriage status*user enrollment status} interaction cause, according to Table 1, "b1maruser" is not statistically significant with respect to 0.05 type I error rate. To come up with this, an approximate $F$ test using the delta method is applied to test that both "b1maruser" and "b2maruser" are simultaneously equal zero. The corresponding P-Value is found to be 0.016, which proves the significance of overall contribution marriage and user-enrollment interaction.

Although our approach works well in terms of prediction precision and detection of significantly contributing patient factors, further improvements can be made in some practical aspects. The most interesting one as discussed by VA health professional is to include some healthcare-connected variables like ICD, treating specialty, or principal diagnosis into the final regression equation [20]. Because it is natural to think that, no matter in which city the patient was admitted, the risk of being readmitted is dissimilar for different disease types. To address this issue, we analyze the data set more carefully and find that the number of levels (or classes) under such features is rather big so putting them into the model can cause it to be over-specified with ambiguous statistical test results [21]. This happens since the required degrees of freedom to fully estimate say ICD is near 85, which means that there should be at least 850-1275 failures (readmissions) existed in the dataset. But the total number of readmissions in the data is only 467. Possible solutions here are to collapse the categories in a meaningful way with some clinical inputs and/or collect more data records in which we are currently working.

*B. Predicting risk of readmission*

In this part, we present a method to actually predict the risk of patient readmission when patient characteristics are inputted. This can enable health providers identify the patients with high chances of non-compliance with discharging instructions, thus targeted post-care strategies such as timely reminders may be developed to effectively reduce further rehospitalization [22]. Assume that readmission risk for $i^{th}$ patient is of interest given patient factors $\mathbf{Z}_i$. A natural point prediction of (2) is obtained by $\lambda(\hat{\beta}, \hat{\gamma}_i)$ where $\hat{\beta}$ is the maximum likelihood estimate of $\beta$ (Table 1) and $\hat{\gamma}_i$ (equivalently $\hat{u}_i$) is the empirical Bayes estimate of $u_i = argmin\{-log[p(y_i|Z_i, \varphi, u_i)q(u_i|\xi)]\}$ with $\hat{\varphi} = [\hat{\beta}, \hat{\omega}]$ and $\hat{\xi} = \hat{\sigma}_u^2$ set equal to their optimal estimates. Also an approximate prediction variance matrix for $(\hat{\beta}, \hat{\gamma}_i)$ is given by

$$\mathbf{V} = \begin{pmatrix} \hat{\mathbf{H}}^{-1} & \hat{\mathbf{H}}^{-1}\left(\frac{\partial \hat{u}_i}{\partial \theta}\right)' \\ \left(\frac{\partial \hat{u}_i}{\partial \theta}\right)\hat{\mathbf{H}}^{-1} & \hat{\Gamma}^{-1} + \left(\frac{\partial \hat{u}_i}{\partial \theta}\right)\hat{\mathbf{H}}^{-1}\left(\frac{\partial \hat{u}_i}{\partial \theta}\right)' \end{pmatrix} \quad (5)$$

where $\theta = [\varphi, \xi]$ i is the vector of unknown parameters, $\hat{H}$ is the approximate Hessian matrix pulled out from optimization for $\hat{\theta}$, $\hat{\Gamma}$ is the approximate Hessian matrix from the optimization for $\hat{u}_i$, and $(\partial \hat{u}_i / \partial \theta)$ is the derivative of $\hat{u}_i$ with respect to $\theta$, evaluated at $(\hat{\theta}, \hat{u}_i)$ [23]. The approximate variance matrix for $\hat{\theta}$ is the inverse Hessian matrix evaluated at $\hat{\theta}$, and that for $\hat{u}_i$ is an approximation to the conditional mean squared error of prediction described in [24]. Based on this approach, we also compute empirical Bayes estimate of the baseline readmission hazard, pdf of readmission time, and survivor function for all patients but for sake of brevity these are not presented here.

Following the above approach, the predicted readmission risk and other related statistics for the patient cohort are summarized in Table 2. Note that in the table, 'Pctl' stands for percentile.

Table 2: Summarized Statistics for Predicted Risk of Readmission

| Min | Max | Med | Mean | Std Dev | 1st Pctl | 5th Pctl | 90th Pctl | 95th Pctl | 99th Pctl |
|---|---|---|---|---|---|---|---|---|---|
| | | | | | | | | | |

| 2.9 E-6 | 8.0 | .06 | .12 | 0.38 | 3.43 E-6 | .90 6E-4 | .181 | 0.48 | 2.3 |

As depicted, the distribution of predicted hazard has relatively few high values and very right skewed. This happens since the total number of readmissions in the data set is only 467 (or equivalently 9.02%). Also may be of interest to see that the hazards, unlike pdf and survivor function, can be greater than 1.0 with no upper bound but it cannot be less than zero. One principal advantage of providing readmission risk predictions is for health providers in a way that enable them to classify patients to "high" and "low" risk groups based on hazard estimates in above table. And then lots of follow-up interventions can be advised to improve rate of unnecessary readmissions.

We finish by reminding the fact that this study, unlike classification-based methods such as logistic regression, centers on timing of events and develops a risk prediction model for hazard rate of readmission in case of repeated-measured responses and random effect covariates. So we feel that doing comparisons between our proposal and those classifiers is although possible but, because of the three main reasons mentioned in section 2.1, may not very precise.

IV. CONCLUSION

In this paper, we formulate an analytical approach based on hierarchical mixed-effect models to systematically reduce the number of avoidable readmissions mainly caused by patient non-compliances to medication instruction. Our proposal has the capability of capturing both patient and population based variations of hospital readmissions. The novelty of our method is to directly incorporating patients' history of readmissions into modeling framework along with other demographic and clinical characteristics. We also verify the effectiveness of the proposed approach based on real dataset from four facilities in the State of Michigan.

Some contributions made in this paper are (i) applying stepwise variable selection in mixed-effect framework and (ii) extending the (normal) random frailty model for Weibull hazard function with patient factors incorporated. We are also working on ways to allow the healthcare-connected features, such as principal diagnosis, be included in the final regression equation in the way to improve the practical aspects of our approach [25,26].

Some research directions can be sought by testing different variable selection techniques such as LASSO or Nonnegative Garrote in mixed-effects models for better subset regressions. Also in presence of high right censored data, it is interesting to consider some health care cost measures from which it may be possible to statistically estimate the mean population cost instead of mean survival times.

ACKNOWLEDGMENT

The authors wish to thank Chief of Systems Redesign in Detroit VA Medical Center Mrs. Susan Yu for her valuable collaboration in this research and providing the patient non-compliances data. This study was supported by VA Center for Applied System Engineering under grant VA251-P-1013 "Patient Discharging Error and Re-admission Reduction".